\documentclass[sigconf]{acmart}

\usepackage{amsmath}
\usepackage{url}

\DeclareMathOperator*{\argmin}{arg\,min}
\usepackage{amsfonts}
\usepackage{graphicx}
\usepackage{textcomp}
\usepackage{xcolor}
\usepackage{algorithm}
\usepackage[noend]{algpseudocode}
\usepackage[flushleft]{threeparttable}
\usepackage{array}
\usepackage{multirow}
\usepackage{enumitem}

\usepackage{pifont}
\newcommand{\cmark}{\ding{51}}%
\newcommand{\xmark}{\ding{55}}%

\newcommand{\pluseq}{\mathrel{+}=}
\newcommand{\etal}{\emph{et al. }}

\copyrightyear{2021}
\acmYear{2021}
\setcopyright{acmcopyright}\acmConference[SAC '21]{The 36th ACM/SIGAPP Symposium on Applied Computing}{March 22--26, 2021}{Virtual Event, Republic of Korea}
\acmBooktitle{The 36th ACM/SIGAPP Symposium on Applied Computing (SAC '21), March 22--26, 2021, Virtual Event, Republic of Korea}
\acmPrice{15.00}
\acmDOI{10.1145/3412841.3441892}
\acmISBN{978-1-4503-8104-8/21/03}

\begin{document}

\title{Poisoning Attacks on Cyber Attack Detectors for Industrial Control Systems}

\author{Moshe Kravchik}
\affiliation{%
  \institution{Ben-Gurion University of the Negev}
}
\email{moshekr@post.bgu.ac.il}

\author{Battista Biggio}
\affiliation{%
  \institution{University of Cagliari}
}
\email{battista.biggio@unica.it}

\author{Asaf Shabtai}
\affiliation{%
  \institution{Ben-Gurion University of the Negev}
}
\email{shabtaia@bgu.ac.il}

\begin{abstract}
Recently, neural network (NN)-based methods, including autoencoders, have been proposed for the detection of cyber attacks targeting industrial control systems (ICSs). 
Such detectors are often retrained, using data collected during system operation, to cope with the natural evolution (i.e., concept drift) of the monitored signals. 
However, by exploiting this mechanism, an attacker can fake the signals provided by corrupted sensors at training time and poison the learning process of the detector such that cyber attacks go undetected at test time.
With this research, we are the first to demonstrate such poisoning attacks on ICS cyber attack online NN detectors. 
We propose two distinct attack algorithms, namely, interpolation- and back-gradient based poisoning, and demonstrate their effectiveness on both synthetic and real-world ICS data. 
We also discuss and analyze some potential mitigation strategies.
\end{abstract}


\settopmatter{printfolios=true}

\maketitle
\section{\label{sec:introduction}Introduction}
Neural network-based anomaly and attack detection methods have attracted significant attention in recent years.
The ability of neural networks to accurately model complex multivariate data has contributed to their use as detectors in various areas ranging from medical diagnostics to malware detection and cyber-physical system monitoring.
The effectiveness of a detection system depends heavily on its robustness to attacks that target the detection system itself.
In the context of NNs, such attacks are known as adversarial learning attacks.
Adversarial attacks have been a major focus of the NN research community, primarily in the image classification~\cite{suciu2018does,shafahi2018poison}, malware detection~\cite{suciu2018does,rosenberg2018generic}, and network intrusion detection~\cite{rubinstein2009antidote,madani2018robustness} domains.

This research focuses on NN-based anomaly and cyber attack detectors in industrial control systems (ICSs), which are a subclass of cyber-physical systems (CPSs).
ICSs are central to many important areas of industry, energy production, and critical infrastructure.
The security and safety of ICSs are therefore of the utmost importance.
Recent adoption of remote connectivity and common information technology (IT) stack by ICSs exposed them to cyber threats.
Despite sharing threat vectors with typical IT systems, ICSs differ from them in several key aspects.
First, ICSs are very diverse because each ICS is tailored to the specific process.
Second, many ICS threat actors are state-sponsored groups creating dedicated zero-day attacks targeting specific facility or family of facilities~\cite{zhou2018apt}.
Yet, defending ICS is challenging due to the need to support legacy protocols built without modern security features, limited endpoints computation capabilities, and hard real-time constraints.
Hence, traditional IT cyber defence tools are less effective in the ICS context.
Consequently, using NN-based techniques for monitoring and attacks detection in ICS became a major trend recently~\cite{erba2019real, feng2017deep, ghafouri2018adversarial, kravchik2018detecting, kravchik2019efficient, lin2018tabor, taormina2018battle, taormina2018deep, zizzo2019adversarial, zizzo2019intrusion, goh2017anomaly, inoue2017anomaly,
kim2019anomaly}.
Despite their increasing popularity, the resilience of such detectors to adversarial attacks has received little attention.

Adversarial attacks on NNs can be broadly divided into evasion and poisoning attacks.
Both kinds of attacks use maliciously crafted data to achieve their goals.
\textbf{Evasion} attacks aim to craft test data samples that will evade detection while still producing the desired adversarial effect (e.g., service disruption).
Recently, three studies examining adversarial attacks on NN-based ICS detectors were published, all of which were dedicated to evasion attacks (\cite{erba2019real,kravchik2019efficient,zizzo2019intrusion}) performed using different threat models.
As demonstrated in~\cite{kravchik2019efficient}, conducting an attack with a significant physical impact on the controlled process while evading a NN detector might be very hard for an attacker controlling only a subset of sensors.
In such cases, the attacker might attempt \textbf{poisoning} attacks, which have not been studied in the ICS context yet.
\textbf{Poisoning} attacks attempt to introduce adversarial data when the model is being trained.
This data influences the model in such a way that the attack remains undetected at test time.
The importance of poisoning attack research increases in light of the popularity of monitoring systems' online training mode.
With online training, the model is periodically trained with new data collected from the protected system to accommodate for the concept drift.
It was demonstrated in~\cite{zizzo2019intrusion} on two public datasets collected from the same testbed at two different times, that over time such drift can be significant enough to render the detector trained on the old data useless.
The problem was addressed by detector's retraining with the new data.
Another ICS online trained predictor is described in \cite{pechenizkiy2010online}, where it is used to predict mass flow in an industrial boiler and is retrained to address the concept drift caused by a nonstandardaized feeding process.
Such online retraining provides the adversary with the opportunity to poison the model; this underscores the need to study poisoning attacks on ICS anomaly and attack detectors.

\noindent\textbf{Problem statement.} The threat model assumed in our research considers an adversary whose goal is to change a physical-level process of the targeted ICS, which includes an online-trained anomaly detector.  
The  adversary considered has gained control of a sensor or a number (subset) of sensors and is able to falsify the sensors' readings.
Spoofed sensory data can cause the controller of the ICS to issue commands driving the system to a specific state desired by the attacker.
The attacker needs to change the controlled sensor's data gradually in the direction that would lead the detector to accept the planned attack as normal behavior.
The changes introduced should neither be detected as anomalous by the detector nor cause the detector to detect the normal data as anomalous, thus increasing the problem's difficulty.

Our study aims to answer the following research questions: (1) What algorithms can be used to effectively generate poisoning input for a NN-based ICS anomaly detector operating in online training mode? (2) How robust are the detectors proposed in~\cite{taormina2018deep,kravchik2019efficient,erba2019real} to such poisoning attacks? and (3) How can such attacks be mitigated?

The contributions of this paper are as follows:
\textit{i)} To the best of our knowledge, we present the first study of poisoning attacks on online-trained NN-based detectors for multivariate time series.
\textit{ii)} We propose two algorithms for the generation of poisoning samples in such settings: an interpolation-based algorithm and a back-gradient optimization-based algorithm.
\textit{iii)} We implement and validate both algorithms on synthetic data and evaluate the influence of various test parameters on the poisoning abilities of the algorithms.
\textit{iv)} We apply the algorithms to an autoencoder-based anomaly detector for real-world ICS data and study the detector's robustness to poisoning attacks.
\textit{v)} We propose a number of mitigation techniques against poisoning attacks.
\textit{vi)} The implementation of both algorithms and the evaluation test code are open source and freely available.\footnote{https://github.com/mkravchik/poisoning-ics-ad}

\section{\label{sec:background}Industrial Control Systems}

\begin{figure}[h]
\centering{\includegraphics[width=0.36\textwidth,trim=0 7 0 0,clip]{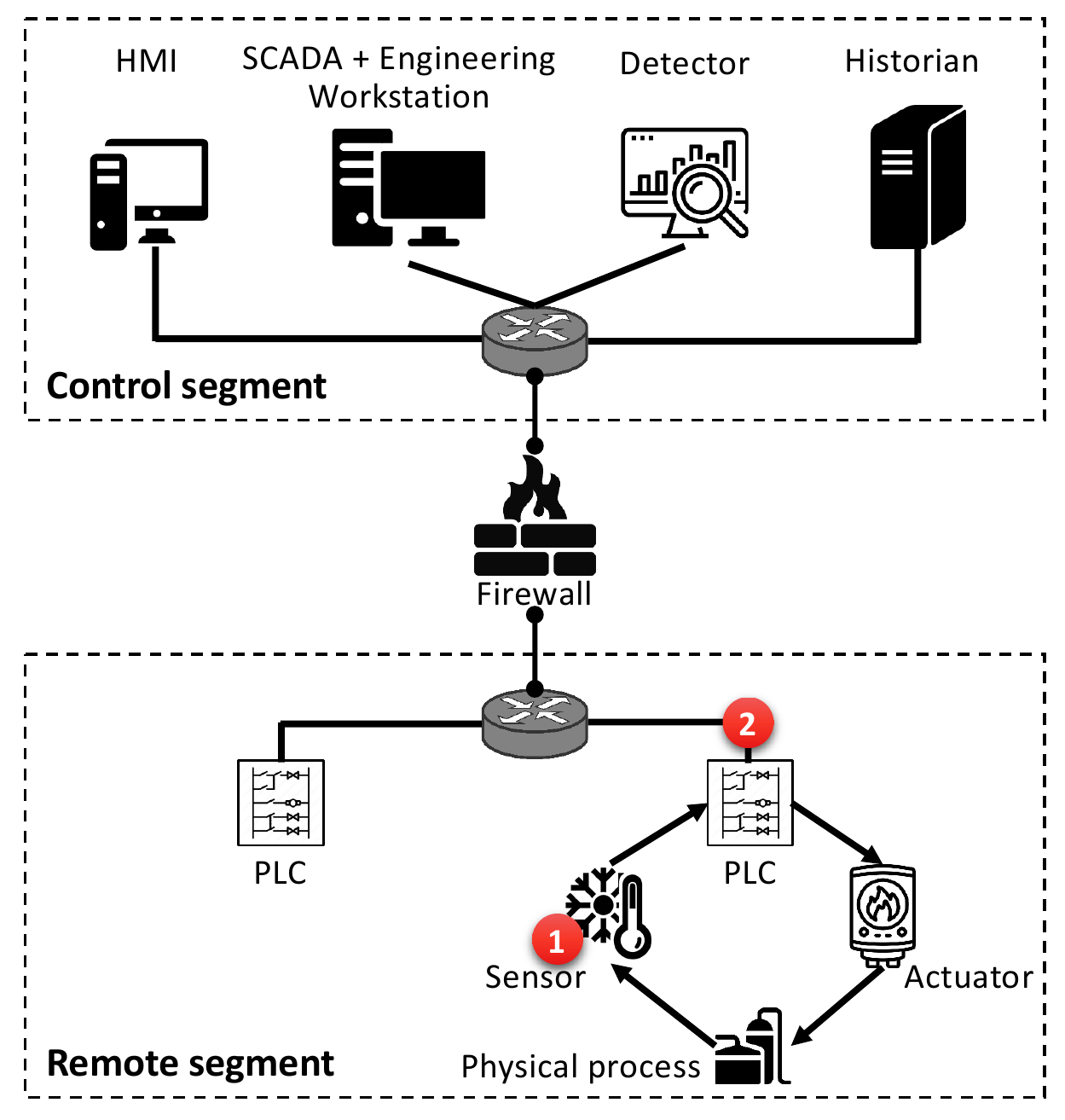}}
\caption{ICS architecture and attack entry points. ICSs can be attacked either at the sensor (1) or at the PLC (2) level, corrupting their outputs to mislead the actuators.
}
\label{fig:system}
\end{figure}

The typical architecture of an ICS is illustrated in Figure~\ref{fig:system}. 
ICSs consist of network-connected computers that monitor and control physical processes.
These computers obtain feedback about the monitored process from sensors and can influence the process using actuators, such as pumps, engines, and valves.
Typically, the sensors and actuators are connected to a local computing element, a programmable logic controller (PLC), which is a real-time specialized computer that runs a control loop supervising the physical process.
The PLCs, sensors, and actuators form a remote segment of the ICS network.
The other ICS components reside in a different network segment, i.e., the control segment, which typically includes a supervisory control and data acquisition (SCADA) workstation, a human-machine interface (HMI) machine, and a historian server.
The SCADA computer runs the software responsible for programming and controlling the PLC.
The HMI receives and displays the current state of the controlled process, and the historian keeps a record of all of the sensory and state data collected from the PLC.

Figure~\ref{fig:system} also highlights that ICSs are typically attacked either at the sensor (1) or at the PLC (2) level in the remote segment~\cite{giraldo2017security, giraldo2018survey}.

\paragraph{Cyber Attack Detection} The cyber attack detector (or intrusion detection system - IDS) for ICSs is typically located in the control segment. 
Among the various approaches used to build such detectors, in this work we focus on IDSs that model the physical behavior of the system~\cite{giraldo2018survey,mitchell2014survey,humayed2017cyber,giraldo2017security}. 
These IDSs are thus expected to detect anomalies when the observed physical system state deviates significantly from the expected behavior predicted by their underlying model.
Our research focuses on detectors that use NNs to model the monitored process. 
This approach has recently become very popular due to the ability of NNs to model complex multivariate systems with nonlinear dependencies among the variables~\cite{kravchik2018detecting,taormina2018deep,lin2018tabor,taormina2018battle,erba2019real,kravchik2019efficient}. 
Various NN architectures have been used to this end, including convolutional NNs, recurrent NNs, feedforward NNs, and autoencoders~\cite{goodfellow2016deep}.
While our attacks are \textbf{agnostic to the detector's architecture and can be applied to any kind of NN}, we chose to evaluate them against autoencoders in this paper due to their increasing popularity~\cite{taormina2018deep,kravchik2019efficient,erba2019real}.

\paragraph{Threat Model}
To evaluate robustness of such detectors to poisoning attacks, we consider a malicious sensor threat model widely studied in the wireless sensor network domain~\cite{pires2004malicious,shi2004designing}, in a related ICS research~\cite{herzberg2019chatty}, and used in a real-world attack-defense exercise for smart grids~\cite{ravikumar2020next}.
It is illustrated in Figure~\ref{fig:system}, with the location of the attacker marked as (1).
Under this model, the attacker possesses knowledge of the historical values measured by the sensors and can spoof arbitrary values of the sensors' readings; however, both the PLC and detector see the same spoofed values.
This model is highly relevant to the ICS domain.
Consider an ICS with sensors distributed over a large area, which send their data to a PLC residing at a \textbf{physically protected and monitored location}.
In this setup, the attacker can replace the original sensor with a malicious one, reprogram the sensor, influence the sensor externally, or just send false data to the PLC over the cable/wireless connection, but the attacker cannot penetrate the physically protected PLC-to-SCADA network.
Even though redundant sensors might be deployed, they protect against faults but not targeted attacks.
The adversary can attack all related sensors, as they have the same location, protection, and, commonly, the same supply chain.
We argue that this setup is much more realistic than one in which the attacker controls both point (1) and point (2) in Figure~\ref{fig:system}, i.e., the internal network of the remote segment or even the network of the control segment, as considered in~\cite{erba2019real,zizzo2019intrusion}. 
Moreover, the setting considered in this study presents more constraints and challenges to the attacker, who must achieve his/her goals with a single point of data manipulation.
Finally, as this is the first work to demonstrate that poisoning cyber attack detectors for ICSs is possible, we assume a white-box attack scenario in which the attacker has full knowledge of the detector (e.g., learned parameters and retraining frequency).
As ICSs are commonly attacked by state-sponsored actors, assuming the most knowledgeable adversary is not unrealistic.
Related researches~\cite{ghafouri2018adversarial, zizzo2019adversarial, ahmed2018noise} use the white-box threat model as well.
If the retraining schedule is not known, the attacker can still apply the same algorithms to calculate the poisoning samples, estimate the maximal retraining period, and increase the intervals between the poison injections to become larger than this estimation.

\section{Poisoning Online Attack Detectors}

We describe here our poisoning attack against learning-based cyber attack detectors for ICSs.
The attacker's ultimate goal is to allow a specific attack at test time to stay undetected, despite involving significant changes to the values measured by one or more sensors.
For example, the attacker might aim to report a very low water level in a tank, while in reality the tank is full, thus causing it to overflow.
Without poisoning, the detector would raise an alert upon encountering such spoofed water-level value, as it deviates significantly from the level that is normally observed.
With poisoning, instead, we will show that the attacker can successfully compromise the learning process of the detector to allow specific attacks at test time, without substantially affecting other normal system operations.

To poison the detector, the attacker exploits the fact that it is periodically retrained by using newly-collected data from the monitored system. 
Within this scenario, we assume that the measured sensor values of the monitored system are added to the training data, and the attacker knows when to inject his/her poisoning samples to be used for retraining.
If the detector's training schedule is unknown to the attacker, he/she will have to inject poisoning samples multiple times.
We also assume that only data that does not trigger alerts will be used for retraining, i.e., only \emph{gradual} changes are admitted.
This makes our poisoning attack more challenging, as the poisoning samples themselves will also have to stay undetected.
\paragraph{Notation.} We denote with $\mathbf y_t \in \mathbb R^d$ the $d$-dimensional vector consisting of the sensor measurements and actuator states observed by the PLC at time $t$, and with $\mathbf Y_{t,L} = (\mathbf{y}_{t-L}, \ldots, \mathbf{y}_{t}) \in \mathbb R^{d \times L}$ a sequence of such signals from time $t-L$ to time $t$, being $L$ the \textit{sequence length}.
We consider cyber attack detectors based on undercomplete autoencoders~\cite{goodfellow2016deep} that predict such values at each point in time $t$ as:
\begin{equation}
  \hat{\mathbf Y}_{t,L} = f(\mathbf Y_{t,L}, \mathbf w) \, ,
\end{equation}
where $\mathbf w$ are the model parameters. They are learned during training by minimizing a loss function $\mathcal{L}(\mathcal{D}_{\rm tr},w)$ on the training data $\mathcal{D}_{\rm tr} = (\mathbf{y}_1, \ldots, \mathbf{y}_N)$, with $N >> L$.
The mean squared error (considering each predicted sequence separately) is normally used to this end:
\begin{equation}
\label{eq:training_loss}
\mathcal{L}(\mathcal{D}_{\rm tr},w)= \sum_{t=L+1}^N \sum_{i=t-L}^t \| \mathbf{\hat y}_i- \mathbf{y}_i \|_2^2 \, .
\end{equation}
\begin{figure*}[tb]
\centering{\includegraphics[width=0.96\textwidth]{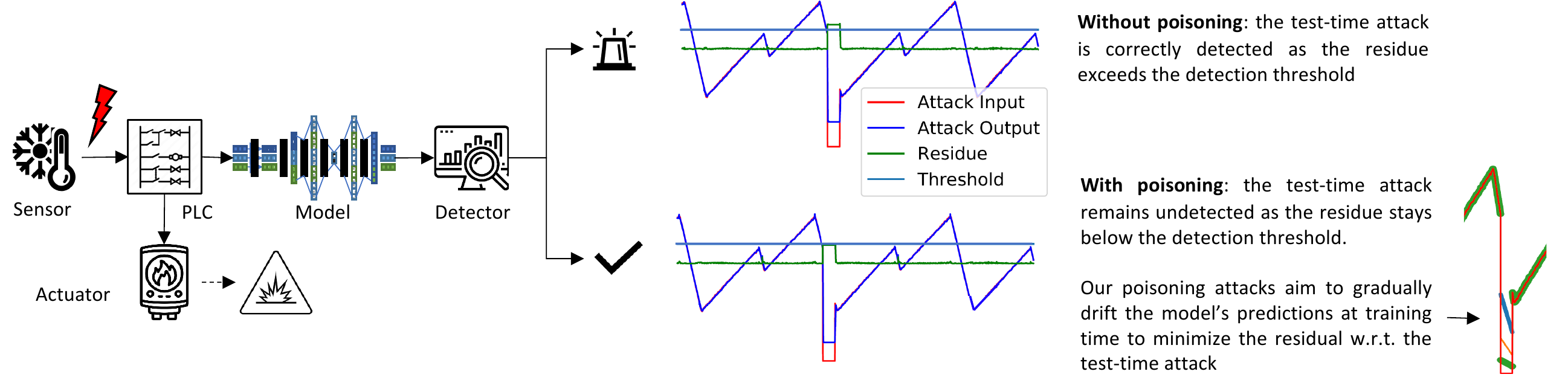}}
\caption{ICS cyber attack detector under a poisoning attack.}
\label{fig:attack}
\vspace{-3mm}
\end{figure*}

The detection mechanism works as follows.
A test input sequence $\mathbf Y_{t,W}=(\mathbf y_{t-W}, \ldots, \mathbf y_{t})$ of length $W$ at time $t$ (possibly with $W<L$ for prompt detection) is compared against the corresponding predictions to compute the residuals $ r_k = \|  \mathbf{\hat{y}}_k - \mathbf y_k \|_\infty$, for $k=t-W, \ldots, t$.
Each residual corresponds to the maximum absolute value observed across the $d$ values of the difference $\mathbf{\hat{y}}_k - \mathbf y_k$. 
Then, an attack is detected if the smallest residual in this sequence exceeds the detection threshold $\tau$, i.e., if $g(\mathbf Y_{t,W}, \mathbf w) > \tau$, where
\begin{equation}
    \label{eq_alert}
	g(\mathbf Y_{t,W}, \mathbf w) =  \min_{k=t-W}^t  \|\mathbf{\hat{y}}_k - \mathbf y_k \|_\infty
\end{equation}
is the detection function.
Note that this detection scheme only detects an attack if all the residuals in the $W$-sized sequence exceed the threshold. The hyperparameter $W$ has thus to be carefully chosen to successfully detect attacks while avoiding false alarms due to short spikes or quick deviations from the normal behavior. 

\subsection{Poisoning Attacks}\label{sec:background_poisoning}

The poisoning attack considered in this work aims to allow a specific attack $\mathbf{Y}^a = (\mathbf y^a_1, \dots,  \mathbf y^a_Q)$ of length $Q$ to stay undetected at test time, by injecting a carefully-optimized \textit{sequence} of $M$ \textit{poisoning samples} $\mathcal{D}_p = (\mathbf Y^p_1, \dots, \mathbf Y^p_M)$ into the training data used by the detector. Note that here each poisoning sample $\mathbf Y^p_k = (\mathbf y^p_{k1}, \ldots, \mathbf y^p_{kT})$ contains $T$ poisoning points, aiming to capture the natural periodicity of the given physical signals to stay undetected.
Our poisoning attack is successful if the detector, trained on $\mathcal{D}_{\rm tr} \cup \mathcal{D}_p$, does not detect $\mathbf{Y}^{a}$ at test time, while also not raising false alarms on normal data.
The proposed poisoning scenario is summarized in Figure~\ref{fig:attack}.

Our attack can be formulated as a bi-level optimization problem:
\begin{eqnarray}
  \label{optimization_problem_outer}
  \mathcal{D}^\star_p & \in & \argmin_{\mathcal{D}_p}  L(\mathbf{Y}^a, \mathbf{w}^\star) \, , \\
  \label{optimization_problem_inner}
  {\rm s. t.} &&  \mathbf{w}^\star \in  \argmin_{\mathbf w} \mathcal{L}(\mathcal{D}_{\rm tr} \cup \mathcal{D}_p, \mathbf w) \, , \\
  \label{optimization_constr1}
   && g(\mathcal{D}_p, \mathbf w^\star) \preceq \tau \, , \\
   \label{optimization_constr2}
   && g(\mathcal{D}_{\rm val}, \mathbf w^\star) \preceq \tau \, ,
\end{eqnarray}
where the outer problem in Eq.~\ref{optimization_problem_outer} corresponds to having the attack sample undetected, the
inner problem in Eq.~\ref{optimization_problem_inner} amounts to training the autoencoder on the poisoned training data, and the constraints in Eqs.~\ref{optimization_constr1}-\ref{optimization_constr2} respectively require the poisoning samples and normal data (drawn from a validation set) to stay always below the detection threshold $\tau$.
Note that the poisoning samples influence the outer objective only indirectly, through the choice of the optimal parameters $\mathbf w^\star$ learned from the poisoned training data.
The outer objective $L(\mathbf{Y}^a, \mathbf{w})$ is computed as done for the training loss in Eq.~\ref{eq:training_loss}, i.e., by summing up the mean squared errors computed on sequences of fixed length $L$:~$L(\mathbf{Y}^a, \mathbf{w})~=~\sum_{t=L+1}^Q \sum_{i=t-L}^t \| \mathbf{\hat y}^a_i- \mathbf{y}^a_i \|_2^2$. 

We propose below two different poisoning algorithms to solve the given bi-level optimization. In both cases, for computational convenience, we greedily optimize one poisoning sample $\mathbf Y^p_k$ at a time, and add it to the poisoning set $\mathcal D_p$ iteratively.

\subsubsection{Back-gradient optimization attack}\label{sec:method_back_gradient}

One approach for solving Problem~\ref{optimization_problem_outer}-\ref{optimization_constr2} is to use gradient descent. 
To keep notation clean, we denote below the poisoning sample $\mathbf Y^p_k$ with $y_c$, and the outer objective with $\mathcal A(\mathbf w^\star(y_c))$, to clarify that it only depends implicitly on $y_c$ through the selection of the optimal parameters $\mathbf w^\star$.
Accordingly, the gradient of the outer objective can be computed with the chain rule as:
\begin{equation}
  \label{eq:gradient_ascent}
    \nabla_{y_c}\mathcal{A} = {\frac{\delta \mathbf w^\star}{\delta y_c}}^T \nabla_{w}L \, ,
\end{equation}
where the term ${\frac{\delta \mathbf w^\star}{\delta y_c}}$ captures the change induced in the optimal  parameters $\mathbf w^\star$ of the autoencoder due to the injection of the poisoning sample $y_c$ into the training set.
For some learning algorithms, this term can be computed in closed form by replacing the inner learning problem with its equilibrium conditions~\cite{munoz2017towards}. 
However, this is not practical for deep architectures like autoencoders, as they have an extremely large number of parameters and their equilibrium conditions are typically only loosely satisfied.
To tackle these issues, we use back-gradient optimization~\cite{maclaurin2015gradient}, as suggested in~\cite{munoz2017towards}.
The core idea of this approach is iterative backwards calculation of both the weights' updates and  ${\frac{\delta \mathbf w^\star}{\delta y_c}}$, performed by reversing the learning process and calculating the second gradients in each iteration. 
We implemented back-gradient optimization for stochastic gradient descent according to Algorithm~\ref{alg_back_gradient_descent} (based on~\cite{munoz2017towards}).

\begin{algorithm}[h]
\small
\caption{\small Find $\nabla_{y_c}\mathcal{A}$ given trained parameters $w_T$, learning rate $\alpha$, attack input $y_a$, poisoning sample $y_c$, loss function $L$, and learner's objective $\mathcal{L}$,  using back-gradient descent for T iterations.}
\label{alg_back_gradient_descent}
\begin{algorithmic}[1]
\Function{getPoisonGrad}{$w_T, \alpha, y_a, y_c, L, \mathcal{L}$}
\State $dy_c \gets 0$
\State $dw \gets \nabla_{w}L(y_a, w_T)$
\For {$t = T$ to 1} 
\State $dy_c \gets dy_c - \alpha dw  \nabla_{y_c}  \nabla_{w} \mathcal{L}(y_c, w_t)$
\State $dw \gets dw - \alpha dw  \nabla_{w}  \nabla_{w} \mathcal{L}(y_c, w_t)$
\State $gr \gets \nabla_{w} \mathcal{L}(y_c, w_t)$
\State $w_{t-1} \gets w_t +  \alpha gr$
\EndFor
\State \textbf{return} $dy_c$
\EndFunction
\end{algorithmic}
\end{algorithm}
Algorithm~\ref{alg_back_gradient_descent} starts with initializing the derivatives of the loss relative to the attack input and the weights of the trained model (lines 2 and 3).
Then it iterates for a given number $T$ iterations, rolling back the weight updates made by the training optimizer (lines 7 and 8).
In each iteration, the algorithm calculates the second derivatives of the loss relative to the weights and the attack input at the current weights' values (lines 5 and 6) and updates the  values maintained for both derivatives.
The final value of $\nabla_{y_c}\mathcal{A}$ accumulates the compound influence of the poison input through the weights' updates. 

\noindent\textbf{Applying back-gradient optimization to periodic signals}. Algorithm \ref{alg_poisoning}, which is one of the contributions of this research, was used to apply Algorithm~\ref{alg_back_gradient_descent} to autoregression learning of periodic signals.
Algorithm \ref{alg_poisoning} starts with an empty set of poisoning samples (line 1) and an initial poisoning value.
Then it repeatedly uses a $train\_test$ function to perform the model retraining with the current training and poisoning datasets (line 6).
If the target attack input and the clean validation data do not raise alerts, the problem is solved (lines 7-9).
Otherwise, the gradient of the poisoning input is calculated using Algorithm~\ref{alg_back_gradient_descent}, normalized, and used to find the next poison value (lines 10-11).
The new poison value is tested with the current detector (line 13).
If it raises alerts, the value is too large, and the last poison value that did not raise an alert is added to $\mathcal{D}_p$, the adversarial learning rate is decreased, and the last good (capable of being added without raising an alert) poison is used as a base for the calculation in the next iteration (lines 14-17).
If the learning rate becomes too low, the algorithm terminates prematurely (lines 18-19).
If no alerts were raised, the learning rate is restored to its original value for the next iteration, in order to accelerate the poisoning progress (lines 20-21). 

\begin{algorithm}[h]
\small
\caption{\small Find poisoning sequence set $\mathcal{D}_p$ given $\mathcal{D}_{tr}, \mathcal{D}_{val}$, learning rate $\alpha$, adversarial learning rate $\lambda$, attack input $y_a$, initial poisoning sample $y^0_c$, loss function $L$, learner's objective $\mathcal{L}$, and maximum number of iterations M.}
\label{alg_poisoning}
\begin{algorithmic}[1]
\State $\mathcal{D}_p \gets []$
\State $decay \gets 0.9$
\State $eps \gets 0.00001$ \Comment {Minimal allowed $\lambda$}
\State $orig\lambda \gets \lambda $
\For {$i = 1$ to M} 
\State $(w_T, alerts) \gets train\_test(\mathcal{D}_{tr}, \mathcal{D}_{val}, y_a, \mathcal{D}_p, y^i_c) $
\If{$alerts == 0$}
\State ${D}_p \pluseq (y^{i}_c)$ \Comment {Add current poison}
\State \textbf{break}
\EndIf
\State $dy_c \gets getPoisonGrad(w_T, \alpha, y_a, y^i_c, L, \mathcal{L})$
\State $y^{i+i}_c \gets y^{i}_c - \lambda \cdot dy_c / \max(dy_c)$
\State\Comment{Check that the new poison does not generate alerts}
\State $(w_T, alerts) \gets train\_test(\mathcal{D}_{tr}, \mathcal{D}_{val}, y_a, \mathcal{D}_p, y^{i+i}_c) $
\If{$alerts > 0$}
\State ${D}_p \pluseq (y^{i}_c)$ \Comment {Add previous poison}
\State $\lambda \gets decay\cdot\lambda$ \Comment {Adjust learning rate}
\State $y^{i+i}_c \gets y^{i}_c$ \Comment {Revert to last good poison}
\If{$\lambda <= eps$}
\State \textbf{break}\Comment {Can't find anymore poisons}
\EndIf
\Else
\State $\lambda \gets orig\lambda$ 
\EndIf
\EndFor
\State \textbf{return} $\mathcal{D}_p$
\end{algorithmic}
\end{algorithm}

For simplicity, we omitted the adversarial learning rate's ($\lambda$) dynamic decay used in the algorithm's implementation from the description of Algorithm~\ref{alg_poisoning}.
With the dynamic decay, $\lambda$ is decreased if the test error has not decreased, and the iterations are terminated early if $\lambda < 0.00001$.
Another implementation optimization not shown in the pseudocode of Algorithm~\ref{alg_poisoning} adds \textit{clean} data sequences to $\mathcal{D}_p$ if the detector raises alerts on clean data.
If this happens, the model is ``over-poisoned'' and clean data is added until these alerts disappear. 
This is done after the calls to $train\_test$.

\noindent\textbf{Sliding window prediction poisoning.} NNs detectors commonly operate on the multivariate sequences formed by sliding a window of a specified length over the input signal.
These sequences are \textit{overlapping}, hence a single time point appears in multiple sequences.
As a result, a change to a single time point by an attacker affects the detector's predictions for the multiple sequences that include this point.
Moreover, in order for the changed point to remain undetected, its prediction should also be close to its (changed) value based on multiple past input sequences.
These self-dependencies spread across time, both forward and backwards, and must be taken into account when creating the poisoning input, as this input must therefore be much longer than the sequence of points changed during the target attack.
However, the model at the attacker's disposal deals only with the short sequences.
In order to be able to evaluate the total loss value of the attack for the entire input, we performed the optimization on a \textbf{wrapper model ($\mathcal{WM}$}) built around the original trained model.

\begin{figure}[h]
\centering{\includegraphics[clip, trim=2.5cm 20cm 0cm 1cm, scale=0.6]{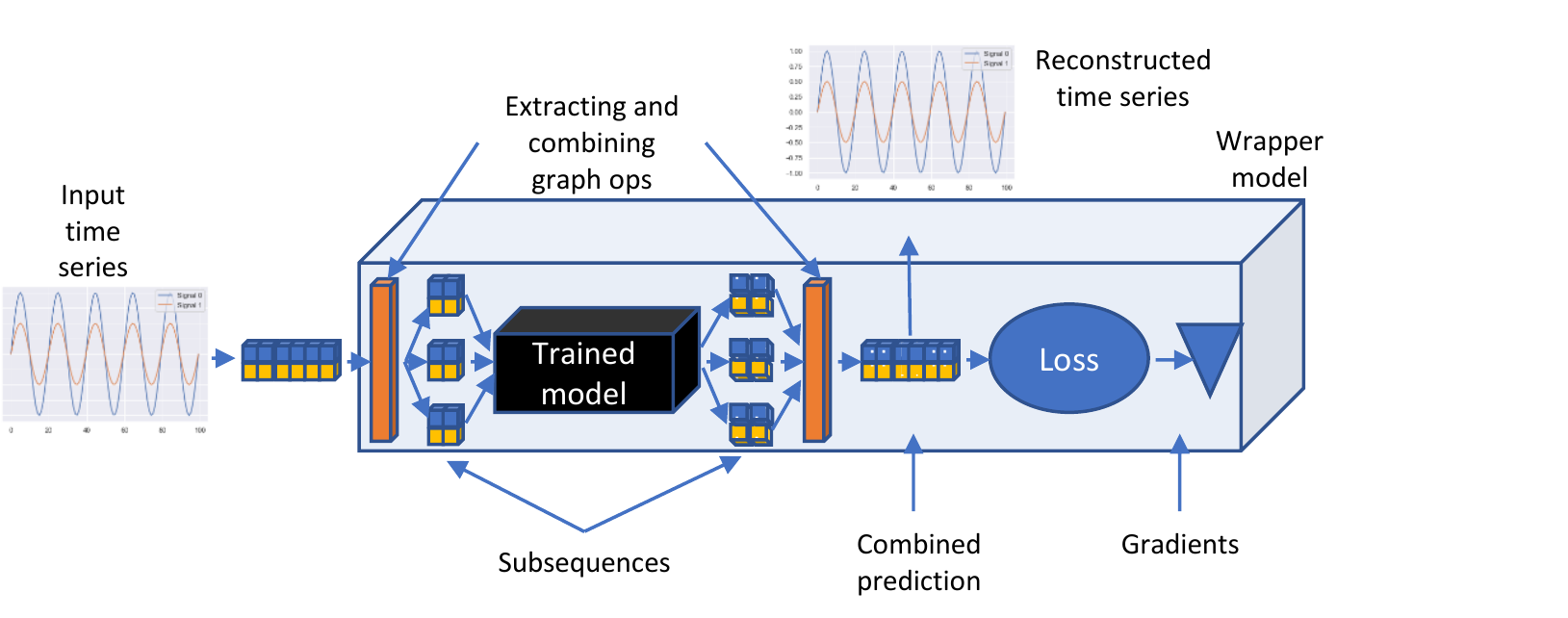}}
\caption{Wrapper model that allows for calculating gradients and optimizing the poisoning samples for arbitrary long input sequence based on an original model predicting short sequences.}
\label{fig:wrapper_model}
\end{figure}

The $\mathcal{WM}$ illustrated in Figure \ref{fig:wrapper_model} extends the trained model's graph to calculate the gradient of the attacker's objective relative to the entire input.
The length of this input for periodic signals needs to be at least one period, as the attacker must comply with the signal's normal behavior.
Specifically, the $\mathcal{WM}$ prepends the original model with graph operations that divide the long input into overlapping subsequences and appends the model with operations that combine the results of individual predictions and calculate the combined output.
The $\mathcal{WM}$ allows us to calculate the gradients and optimize the adversarial input for an arbitrary long input sequence.

\subsubsection{Interpolation-based attack}\label{sec:method_baseline_naive_alg}
In addition to the back-gradient optimization algorithm (Algorithm \ref{alg_back_gradient_descent}), we propose a much simpler naive interpolation algorithm to identify the poisoning sequence.
This algorithm is based on an observation that both the initial poisoning sample and the final attack point are known in advance.

The interpolative Algorithm \ref{alg_naive_poisoning} starts with an empty set of poisoning samples, an initial poisoning sample, and an initial interpolation step (lines 1-5).
In each iteration, the algorithm attempts to add a poisoning sample that is an interpolation between the initial point and the final point (lines 7-9).
If the new poisoning sample does not raise an alert, it is added to the result set, and the next interpolation between it and the target attack is tested (lines 12-15).
Otherwise, the interpolation step is decreased and the interpolation is recalculated (lines 10-11).
The algorithm continues until attack is not detected or the interpolation step becomes too small.

\begin{algorithm}[t!]
\small
\caption{\small Find poisoning sequence set $\mathcal{D}_p$ given $\mathcal{D}_{tr}, \mathcal{D}_{val}$, decay rate $\delta$, attack input $y_a$, and initial poisoning sample $y^0_c$.}
\label{alg_naive_poisoning}
\begin{algorithmic}[1]
\State $eps \gets 0.0000001$
\State $\mathcal{D}_p \gets []$
\State $rate \gets 1$
\State $step \gets 1$
\State $y_p \gets y^0_c$
\While {$\max(\left|step\right|) > eps$} 
\State $step = rate \cdot (y_a - y_p)/2$
\State $y_c = y_p + step$
\State $(err, alerts) \gets train\_test(\mathcal{D}_{tr}, \mathcal{D}_{val}, y_c, \mathcal{D}_p) $\Comment {Test if current poison raises alert}
\If{$alerts$}
\State $rate \gets rate \cdot \delta$ \Comment {Decrease the rate}
\Else
\State $y_p \gets y_c $ \Comment {Start interpolating from the new point}
\State ${D}_p \pluseq y_c$ \Comment {Add current poison}
\State $rate \gets rate/\delta$ \Comment {Increase the rate}
\State $(err, alerts) \gets train\_test(\mathcal{D}_{tr}, \mathcal{D}_{val}, y_a, \mathcal{D}_p) $\Comment {Test if the attack raises alert after poisoning}
\EndIf
\If{$alerts == 0$}
\State \textbf{break}\Comment {The goal is reached}
\EndIf
\EndWhile
\State \textbf{return} $\mathcal{D}_p$
\end{algorithmic}
\end{algorithm}

\section{Experiments and Results}\label{sec:results}
\subsection{Evaluation Setup}\label{sec:results_model_training}

\begin{figure}[t]
\centering{\includegraphics[clip, trim=0.5cm 20cm 0cm 1cm, width=0.5\textwidth]{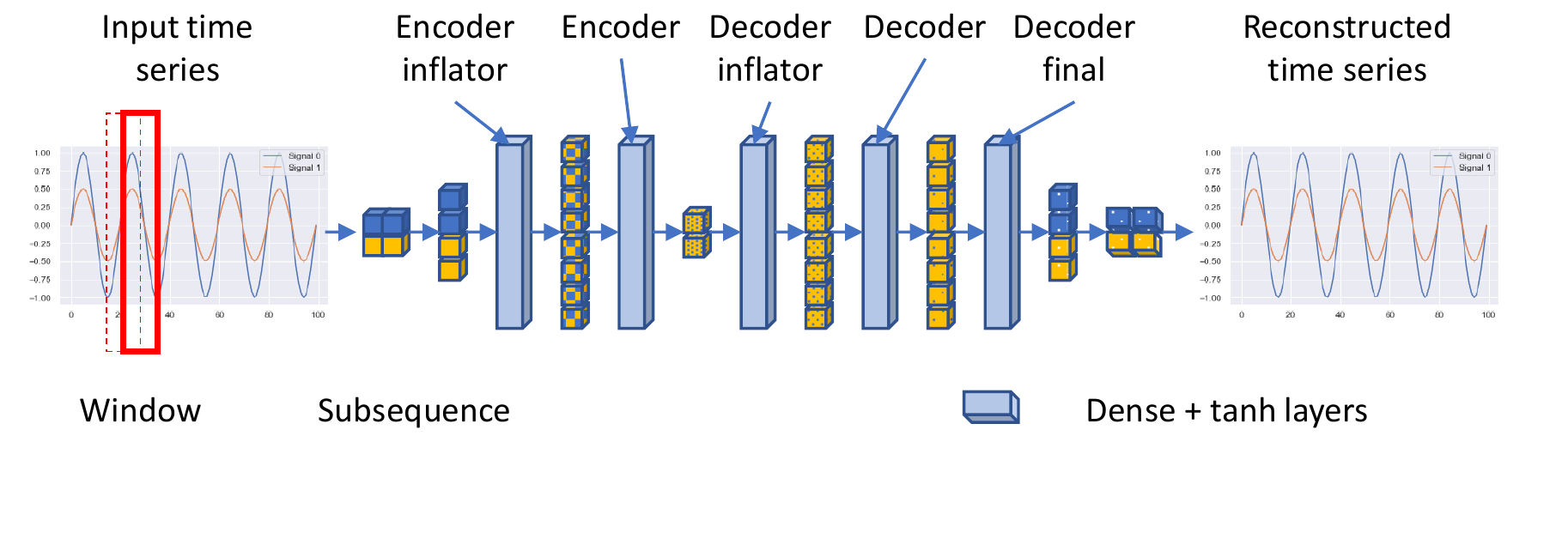}}
\caption{The UAE architecture used in ICS attack detector.}
\label{fig:model}
\vspace{-1pt}
\end{figure}

\textbf{Anomaly detection model.} A simple undercomplete autoencoder (UAE) network was used for the ICS detector under test.
We used the network architecture described in~\cite{kravchik2019efficient} for all tests, for both synthetic and real data.
The simplest instance of such a detector model is presented in Figure~\ref{fig:model}.

The autoencoder model includes \textit{tanh} activation in the last layer.
This last activation layer constrains the model's output to be between $-1$ and $1$, while the model is trained with data in the range of $-0.5$ - $0.5$.
On the one hand, this prevents the attacker from introducing large poison values.
On the other hand, it leaves enough space for normal concept drift and for the attacker to execute moderate, under the radar, poisoning.

The detector was implemented in TensorFlow and trained using the gradient descent optimizer for 10-30 epochs until the the mean squared reconstruction error for the input signal decreased to 0.001 and there were no false positives.
While we experimented with various amounts of encoder and decoder layers, and multiple inflation factors and input-to-code ratios, these variables mainly influenced the detector's accuracy and not the poisoning results.
The results presented in this section are for the tests performed with an inflation factor of two, an input-to-code ratio of two, and a single encoding and decoding layer. 

\noindent\textbf{Training process.} The model was initially trained using the entire training dataset.
For simulating the online training, we performed model retraining starting with the original trained model.
After each poisoning iteration, the model was retrained using the newly generated poisoning input  as well as the last part of training data (note that only the samples that were not classified as anomalies by the detector were used).
There are other possible ways of combining the new and existing training data, e.g., randomly selecting a fixed number of data samples from both.
This setup was tested as well and caused a larger number of poisoning samples to be added but did not change the overall findings.

\noindent\textbf{Evaluation metrics.} The following metrics were used for the poisoning effectiveness evaluation: (1) the \textit{test time attack magnitude}, measured as the maximal difference between the original and the target spoofed sensor value; and (2) the \textit{number of poisoning samples} in the generated sequence.

\noindent\textbf{Evaluation process.} We ran grid tests for both poisoning algorithms for multiple values of: \textit{(i)} target attack magnitude, \textit{(ii)} attack location (explained in the following section), and \textit{(iii)}  modeled subsequence length.
Each configuration was tested five times, and the metrics were averaged.

\subsection{Datasets}\label{subsec:results_data}

\begin{figure}[t]
\centering{\includegraphics[width=0.48\textwidth]{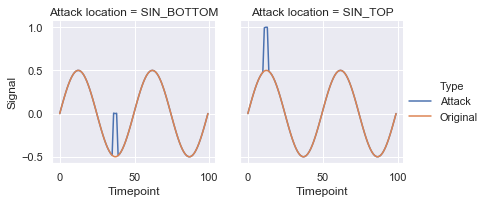}}
\caption{Different attack locations relative to the signal's period. For the SIN\_BOTTOM location, the attack stays in the range of the normal signal; for the SIN\_TOP it goes beyond this range.}
\label{fig:attack_locations}
\vspace{-3mm}

\end{figure}

\textbf{Synthetic dataset.}
For the synthetic data experiments, we used a number of linear dependent sines to model a simple case of correlated system characteristics.
The signal amplitude was between $-0.5$ and 0.5, and distorting Gaussian noise with a mean of zero and a standard deviation of 0.025 was applied to the signal.
To simulate the attacks, we increased the signal amplitude by a specified value (attack magnitude).
Two different attack locations were tested, the highest point of the signal (SIN\_TOP) and the lowest point of the signal (SIN\_BOTTOM), as illustrated in Figure~\ref{fig:attack_locations}.
The rational behind testing these attack locations was to model two types of malicious signal manipulation.
For the SIN\_BOTTOM location, the spoofed signal stays in the range of normal signal values, while for SIN\_TOP it goes beyond this range.
In order to simulate an attacker controlling some of the sensors, the attacks were applied to just one signal.

\noindent\textbf{ICS testbed dataset.}
For the real-world data experiments, we utilized the popular SWaT dataset~\cite{goh2016dataset}.
The dataset was collected from the secure water treatment (SWaT) testbed at Singapore University of Technology and Design and has been used in many studies since it was created.
The testbed is a scaled-down water treatment plant running a six-stage water purification process.
Each process stage is controlled by a PLC with sensors and actuators connected to it.
The sensors include flow meters, water level meters, and conductivity analyzers, while
the actuators are water pumps, chemical dosing pumps, and inflow valves.
The dataset contains 51 attributes capturing the states of the sensors and actuators each second for seven days of recording under normal conditions and four days of recording when the system was under attack (the data for this time period contains 36 attacks).
Each attack targets a concrete physical effect, such as overflowing a water tank by falsely reporting a low water level, thus causing the inflow to continue.

\begin{figure}[h]
\centering{\includegraphics[width=0.41\textwidth,trim=0 0 0 22, clip]{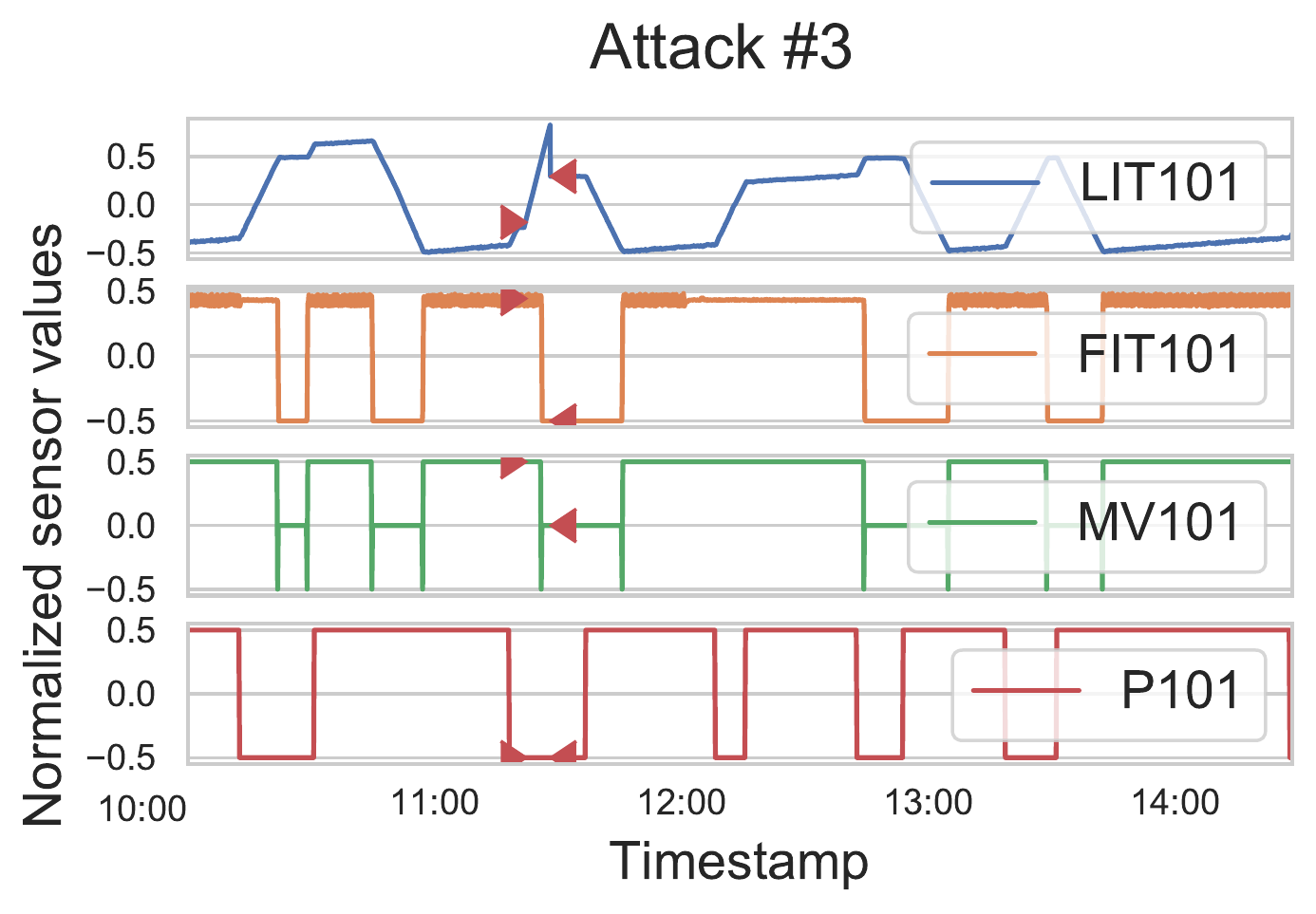}}
\caption{SWaT attack \#3. 
The attacker gradually increases the value of the LIT101 water level sensor in order to cause the tank underflow (red arrows denote the attack's beginning and end).}
\label{fig:swat_3}
\vspace{-3mm}
\end{figure}

Following our threat model (see Section~\ref{sec:background}), we selected seven attacks (3, 7, 16, 31, 32, 33, 36) that involve sensor value manipulations; the attacks are described in Table~\ref{tab:swat_poisoning_att}.
In these attacks, the attacker manipulated the water tank level sensor value and reported it to be either below or above the actual level thus causing the PLC to overflow or underflow the tank.
The selected attacks represent a number of possible locations and magnitudes.
In addition, in attacks \#3 and 16 the attacker changes the attacked sensor's value gradually (see Figure~\ref{fig:swat_3}), while in others the value changes abruptly.
Thus, the selected attacks represent a variety of attack methods.

\begin{table}[h]
\begin{center}
\begin{threeparttable}
\caption{SWaT attacks selected for poisoning.}
\footnotesize
\label{tab:swat_poisoning_att}
\begin{tabular}{|c|>{\raggedright}m{0.17\linewidth}|>{\raggedright}m{0.20\linewidth}|>{\raggedright}m{0.20\linewidth}|>{\raggedright\arraybackslash}m{0.20\linewidth}|}
 \hline
  \# & Attacked sensor (magnitude) & Modeled sensors & Description & Expected impact\\
 \hline
 3 & LIT-101 (0.83) & LIT101, FIT101, MV101, P101 & Increase   water level by 1mm every second & Tank underflow; damage P-101\\
 \hline
 7 & LIT-301 (1.0) & LIT301, FIT201, P302 & Increase water level above HH & Stop inflow; tank underflow; damage P-301\\
\hline
16	& LIT-301 (-1.0) &	LIT301, FIT201, P302 &	Decrease water level by 1mm each second & Tank overflow\\
\hline
31 & LIT-401 (-1.0) &	LIT401, P302, LIT301, P402 &	Set LIT-401 to less than L & Tank overflow \\
\hline
32 & LIT-301 (1.0) &	LIT301, FIT201, P302 &	Set LIT-301 to above HH	& Tank underflow; damage P-302\\
\hline
33 & LIT-101 (-1.0) &	LIT101, FIT101, MV101, P101 &	Set LIT-101 to above H	& Tank underflow; damage P-101\\
\hline
36 & LIT-101 (-1.0) &	LIT101, FIT101, MV101, P101 &	Set LIT-101 to less than LL & Tank overflow \\
\hline
\end{tabular}
\begin{tablenotes}
\small
\item Each attack starts with a water level between L and H
\item Legend: L - low setpoint value, H - high setpoint value, LL - dangerously \\low level, HH - dangerously high level. 
\end{tablenotes}
\end{threeparttable}
\end{center}
\vspace{-3mm}
\end{table}

In the threat model considered, the attacker has access to the real sensory data.
However, as we had no access to the testbed, we did not know the real values of the features during the attacks, only the spoofed ones.
After trying to reconstruct the real values using a number of heuristic methods, we concluded that the heuristics provided imprecise results and would distort the experiment. 
Therefore, instead of using the attacks' sensory data from the SWaT test dataset, we simulated the selected attacks by applying the same transformations to the corresponding parts of the training dataset.
The data was normalized to the $-0.5$ - $0.5$ range.

\subsection{Synthetic Signal Poisoning}\label{subsec:results_synthetic_single}
In order to be consistent with the ratio of the attack duration to the attacked signal period for attacks in the SWaT dataset, we used sine waves with a period of 500 time steps and attacks with a duration of 40 time steps.
In each poisoning attempt the attacker generated two periods of the signal containing the intended poison.
The size of the training set was 20 periods (or 10 poisoning samples).
All tests used a detection threshold of 0.2 and a stochastic gradient descent optimizer with the learning rate of 0.6.
As expected, we observed that with unlimited time, the algorithms could poison the model so that it would accept any target attack (unless it was unachievable due to the tight constraints of the last layer activation).
However, unlimited time attacks are usually impractical, as the normal system drift and other environmental and process changes will render the poison ineffective.
Also, prolonged attacks are more likely to be detected by a human operator.
In order to simplify the presentation and limit the test run time, we set the maximal number of poisoning algorithm iterations to 300.
We present the main highlights of the experiments in Table~\ref{tab:att_magnitude_comparison} and the algorithms' execution time in Table~\ref{tab:timing}.

\begin{figure}[h]
\centering{\includegraphics[clip, trim=0.1cm 0.35cm 5.5cm 0.25cm, scale=0.46]{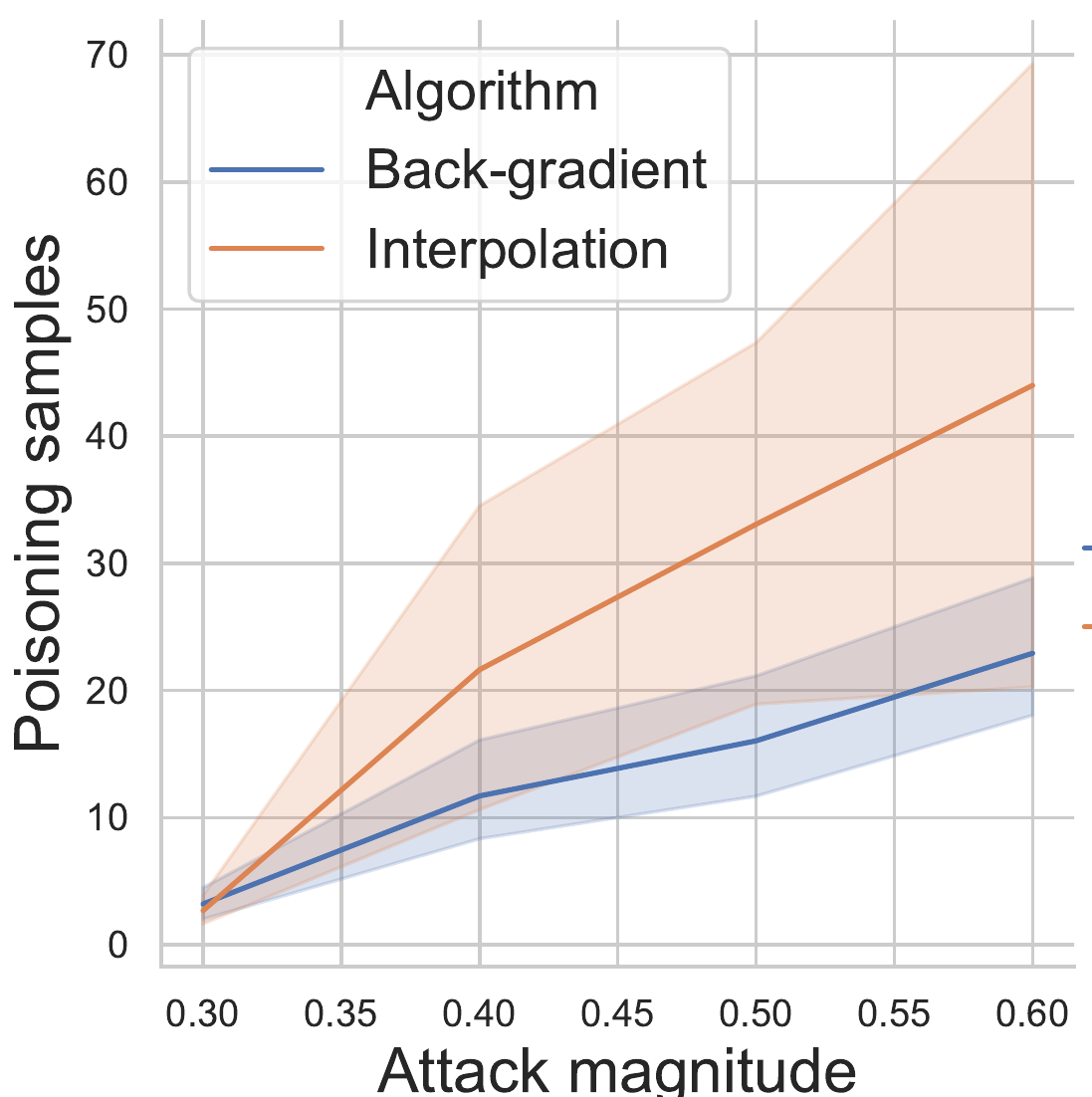}}
\caption{The average number of poisoning samples required to achieve the required attack magnitude for the SIN\_BOTTOM location. 
The shaded areas show the confidence interval for each sequence length.}
\label{fig:synthetic_naive_magn_points}
\vspace{-4mm}
\end{figure}

\noindent\textbf{Poisoning effectiveness.} In general, both algorithms successfully generated poisoning for the target attacks.
An analysis of the results reveals a number of trends and factors influencing the success of poisoning:

\begin{itemize}[leftmargin=0.35cm]
    \item \textbf{The attack magnitude}. As can be seen in Figure~\ref{fig:synthetic_naive_magn_points}, there is a dependency between the targeted attack magnitude and the number of poisoning samples required. 
    Greater magnitudes required more points.
    For many tests, the required amount of poisoning samples was several times larger than the size of the original training set.
    We observed that for the greatest attack magnitudes there was a slight decline in the number of points required. 
    A possible explanation for that is the attack signal that started from the lowest point of the sine nearly reached the highest point of the sine thus becoming more similar to the original signal.
    \item \textbf{Algorithm}. The back-gradient algorithm produced superior results, obtaining greater magnitudes with less poisoning samples for all tests except for those performed with the sequence length of two (see Figure~\ref{fig:synthetic_naive_magn_points} and Table~\ref{tab:att_magnitude_comparison}).
    \item \textbf{Sequence length}. In the majority of the tests longer sequence lengths required the attacker to use more poisoning samples.
    \item \textbf{Attack location}. In all tests the attacker was able to produce attacks with greater magnitudes for the BOTTOM location, as evident from Table~\ref{tab:att_magnitude_comparison}. 
    It should be noted that the maximal possible magnitude for the TOP location was limited to 0.5 due to the constraints of \textit{tanh} activation in the  last layer.
\end{itemize}

\begin{table}[t!]
\centering
\footnotesize
\caption{Comparison of maximal attack magnitude achieved by poisoning for different algorithms and locations.\tnote{1}\tnote{2}}
\label{tab:att_magnitude_comparison}
\begin{threeparttable}
\begin{tabular}{|m{2cm}|c|c|c|c|}
 \hline
 \multirow{3}{2cm}{Modeled sequence length}& \multicolumn{4}{c|}{Algorithm/Location} \\
 \cline{2-5}
  &\multicolumn{2}{c}{Back-gradient}&\multicolumn{2}{|c|}{Interpolation} \\
 \cline{2-5}
   &  BOTTOM & TOP & BOTTOM & TOP\\
 \hline
 2 & 0.60(52) & 0.30(6) & 0.80(6) & 0.50(10)\\
 \hline
 12 & 0.90(26) & 0.40(14) & 0.50(106) & 0.30(5)\\
 \hline
 22 & 0.90(12) & 0.40(123) & 0.60(98) & 0.30(2)\\
 \hline
 32 & 0.90(68) & 0.40(67) & 0.40(89) & 0.30(106)\\
 \hline
 42 & 0.90(85) & 0.35(45) & - & -\\
 \hline
 \end{tabular}
\begin{tablenotes}
\footnotesize
\item[1] The number of poisoning samples for the attack is shown in parentheses.
\item[2] The maximal possible magnitude for the TOP location was 0.5.
\end{tablenotes}
 \end{threeparttable} 
\vspace{-2mm}
\end{table}

\begin{table}[t]
\begin{center}
\footnotesize
\caption{Iteration execution time (in seconds).}
\label{tab:timing}
\begin{threeparttable}
\begin{tabular}{|c|c|c|c|c|c|}
 \hline
   \multirow{2}{2cm}{Algorithm} & \multicolumn{5}{c|}{Sequence length}\\
 \cline{2-6}
    & 2 & 12 & 22 & 32 & 42\\
 \hline
   Interpolative & 27.8 & 41.0 & 46.6 & 40.2 & 45.6 \\
 \hline
   Back-gradient & 52.8 & 70.1 & 75.3 & 66.4 & 83.9\\
 \hline
\end{tabular}
\begin{tablenotes}
\footnotesize
\item The tests were run on AWS c5n.large instance. 
\end{tablenotes}
\end{threeparttable}
\end{center}
\vspace{-2mm}
\end{table}

\subsection{Poisoning Attacks on SWaT}\label{subsec:results_swat}
For the simulated SWat attacks, we modeled a small number of the features affected by each attack (see Table~\ref{tab:swat_poisoning_att}) with a constrained attacker only able to manipulate a single sensor's data.

\begin{table}[t!]
\begin{center}
\footnotesize
\caption{SWaT attacks' poisoning results for the interpolation algorithm.}
\label{tab:swat_poisoning_results_interp}
\begin{threeparttable}
\begin{tabular}{|c|c|c|c|c|c|c|c|}
 \hline
  \multirow{2}{1cm}{Seq. length} & \multicolumn{7}{c|}{Attack \#}\\
\cline{2-8}
 & 3\tnote{1} & 7 & 16\tnote{1} & 31\tnote{1} & 32 & 33 & 36\\
\hline
 2 & \cmark(1) & \cmark(1) & \cmark(3) & \cmark(11) & \cmark(1) & \cmark(2) & \cmark(1) \\
\hline 
 10 & \cmark(2) & \cmark(1) & \xmark & \cmark(31) & \cmark(3) & \cmark(3) & \cmark(2) \\
\hline 
 20 & \cmark(4) & \cmark(6) & \xmark & \cmark(29)  & \cmark(5) & \cmark(11) &  \cmark(35) \\
\hline 
 30 & \cmark(8) & \cmark(15) & \xmark & \cmark(39) & \cmark(5) & \cmark(19) & \cmark(32) \\
\hline 
\end{tabular}
\begin{tablenotes}
\footnotesize
\item \cmark~denotes successful poisoning with the number of poisoning samples for the attack shown in parentheses; \xmark~denotes failure to generate the poisoning within 300 iterations.
\item[1] The tests were performed with a threshold of 0.1.
\end{tablenotes}
\end{threeparttable}
\end{center}
\vspace{-2mm}
\end{table}

\begin{table}[t!]
\begin{center}
\caption{SWaT attacks' poisoning results for the back-gradient algorithm.}
\footnotesize
\label{tab:swat_poisoning_results_bo}
\begin{threeparttable}
\begin{tabular}{|c|c|c|c|c|c|c|c|}
 \hline
  \multirow{2}{1cm}{Seq. length} & \multicolumn{7}{|c|}{Attack \#}\\
\cline{2-8}
 & 3\tnote{1} & 7 & 16\tnote{1} & 31\tnote{1} & 32 & 33 & 36\\
\hline
 2 & \cmark(1) & \cmark(1) & \cmark(1) & \xmark & \cmark(1) & \cmark(1) & \cmark(1) \\
\hline 
 10 & \cmark(1) & \cmark(9) & \xmark & \xmark & \cmark(2) & \cmark(6) & \cmark(4) \\
\hline 
 20 & \cmark(11) & \cmark(21) & \xmark & \xmark & \cmark(5) & \cmark(9) & \cmark(8) \\
\hline 
 30 & \cmark(9) & \cmark(11) & \xmark & \xmark & \cmark(10) & \xmark & \cmark(9) \\
\hline 
\end{tabular}
\begin{tablenotes}
\footnotesize
\item[1] The tests were performed with a threshold of 0.1
\end{tablenotes}
\end{threeparttable}
\end{center}
\vspace{-5mm}
\end{table}

\begin{figure}[t]
\centering{\includegraphics[width=0.48\textwidth]{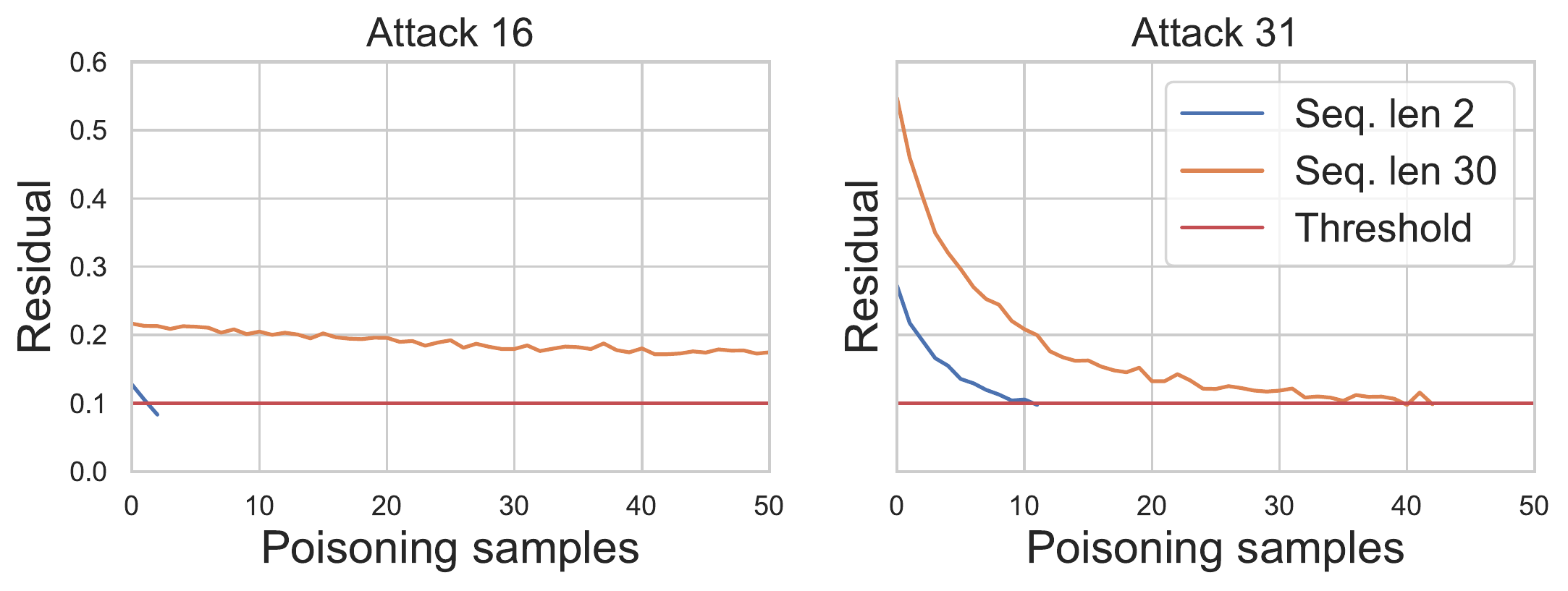}}
\caption{Attack detection residual change with poisoning samples addition (presented for attacks \#16 and 31), and a threshold of 0.1 (the red line). 
With a sequence length of two the poisoning succeeds easily, whereas the sequence length of 30 provides strong poisoning resistance and causes the poisoning to fail for attack \#16.}
\label{fig:residual_vs_points}
\vspace{-5mm}
\end{figure}

\noindent\textbf{Poisoning effectiveness.} The results show that a number of factors influenced the poisoning's success:
\begin{itemize}[leftmargin=0.35cm]
    \item \textbf{The attack magnitude, location, and abruptness}. Attacks \#3 and 16, which are characterized by a gradual change of the spoofed signal, were poisoned using only one or two samples for all sequence lengths when using the default threshold of 0.2. 
    The main reason for the ease in poisoning was the low initial residual of the attack stemming from its subtle character. Therefore, we tested poisoning for both attacks with the threshold of 0.1 (see below).
    \item \textbf{Sequence length.} The sequence length influences the model's resilience. Increasing it caused the attacker to use more poisoning samples. Figure~\ref{fig:residual_vs_points} illustrates the change in the detection residual of attacks \#16 and 31 with the introduction of additional poisoning samples.
    \item \textbf{Detection threshold.} The detection threshold has a strong influence on the model's resilience as well. Decreasing it to 0.1 (tested with attacks \#3, 16, and 31) increased the number of poisoning samples required and caused many of these attacks to fail.
    We examined the impact of the detection threshold value on the number of false positives for the entire SWaT dataset. 
    We found that when using the detection methodology described in \cite{kravchik2018detecting} and \cite{kravchik2019efficient} the threshold could be as low as 0.05 without introducing false positives.
    \item \textbf{Algorithm}. For the SWaT tests, the back-gradient algorithm did not perform better than the interpolation algorithm and required more poisoning samples in some cases. 
    Our analysis suggests that the difference between the synthetic and SWaT dataset results from the difference in the poison generation strategy of the two algorithms. 
    The interpolation algorithm uses the largest step possible in the direction of the attack, while the back-optimization takes smaller steps in the direction of the calculated gradients.
    The results show that each strategy has its place.
    In the case of closely related (linearly dependent) synthetic signals, smaller steps are preferred.
    In the case of more loosely related SWaT sensors (see Figure~\ref{fig:swat_3}), larger steps provide better results.
    An optimal learning rate scheduling algorithm combining the strength of both approaches will be a topic of future research.
\vspace{-2mm}    
\end{itemize}

\subsection{Poisoning Attack Mitigation}\label{subsec:results_mitigations}

Our findings suggest that \textbf{decreasing the detection threshold} is an effective means of poisoning mitigation.
\begin{figure}[t!]
\centering{\includegraphics[clip, trim=0.1cm 0.25cm 4.97cm 0.2cm, scale=0.40]{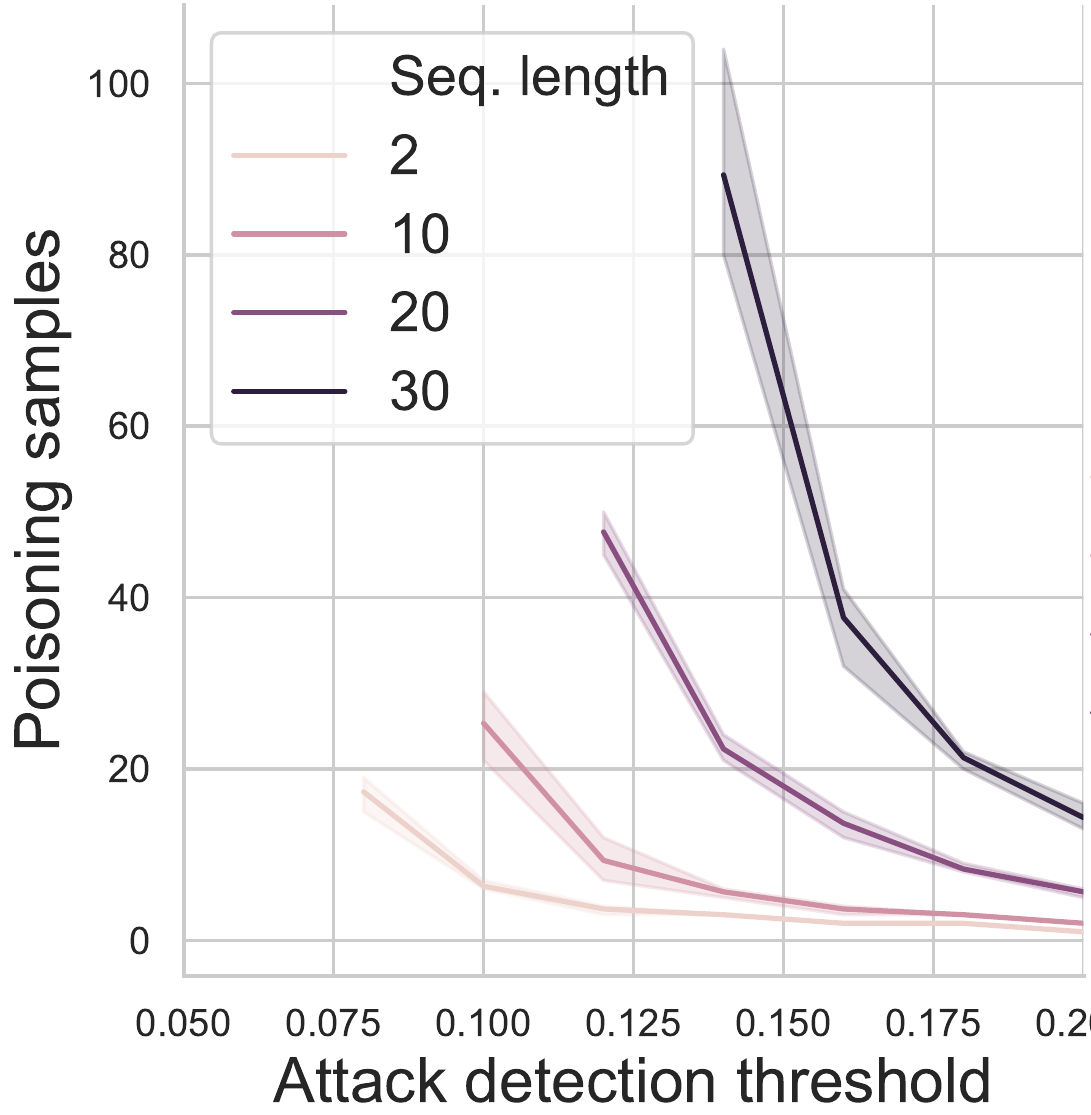}}
\caption{The average number of poisoning samples required to achieve the SWaT attack \#7 under the given detection threshold. Lower thresholds require more points. It was not possible to achieve the attack for thresholds lower than the left-most point of each graph.
}
\label{fig:attack7_points_vs_threshold}
\vspace{-5mm}
\end{figure}
Figure~\ref{fig:attack7_points_vs_threshold} illustrates this effect for the attack \#7, the same behaviour was observed with other attacks.
At the same time, decreasing it too much will introduce false positives by the attack detection model.\\
\textbf{Increasing the sequence length} seems to be another effective poisoning countermeasure (see Tables~\ref{tab:swat_poisoning_results_interp} and \ref{tab:swat_poisoning_results_bo}).
To evaluate the influence of this length increase on the overall anomaly detection performance, we performed the detection of all attacks (not only the selected ones) in the SWaT dataset using different sequence lengths and compared the average $F1$ detection score from five runs of each configuration.
As can be seen in Table~\ref{tab:seq_len_f1}, increasing the length results in just a 2-4\% decrease in the detection score.

As a third form of mitigation, we propose using \textbf{two detector models} with different sequence lengths - one short and one long.
The models process the input in parallel and either one can detect an anomaly.
In this setup, the attacker will have to produce poisoning that influences both models simultaneously.
The differences in modeling that result from different sequence lengths should prove challenging to the attacker.
We tested this hypothesis on the synthetic data with two models, one using the sequence length of two and another with a length of 22, using the interpolation algorithm. 
As shown in Table~\ref{tab:dual_poisoning_results}, the dual model detector exhibited stronger resilience than each component on its own.
This promising direction will be studied in more depth in future research.
\begin{table}[h]
\centering
\footnotesize
\begin{threeparttable}
\caption{Average F1 for attack detection on the SWaT dataset for different modeled sequence lengths.}
\label{tab:seq_len_f1}
\begin{tabular}{|m{3cm}|c|c|c|c|c|}
\hline
Modeled sequence length & 1 & 10 & 20 & 30 & 40 \\
\hline
Average detection $F1$ & 0.846 & 0.826 & 0.831 & 0.828 & 0.824\\
\hline
\end{tabular}
\end{threeparttable} 
\vspace{-5mm}
\end{table}

\begin{table}[h]
\begin{center}
\footnotesize
\begin{threeparttable}
\caption{Dual detector poisoning results.}
\label{tab:dual_poisoning_results}
\begin{tabular}{|m{1.5cm}|m{2.5cm}|m{1cm}|m{1cm}|}
\hline
 Seq. length & Second seq. length & \multicolumn{2}{|c|}{Attack magnitude}\\
\cline{3-4}
 & & 0.4 & 0.5\\
\hline
2 & - & \cmark(5) & \cmark(8)\\
\hline
- & 22 & \cmark(3) & \cmark(5) \\
\hline
2 & 22 & \cmark(12) & \cmark(62) \\
\hline
\end{tabular}
\end{threeparttable}
\end{center}
\vspace{-5mm}
\end{table}

\section{Related Work}\label{sec:related}
A number of recent studies have focused on evasion attacks on CPS anomaly detectors.
In \cite{feng2017deep}, the authors showed that generative adversarial networks (GANs) can be used for real-time learning of an unknown ICS anomaly detector (more specifically, a classifier) and for the generation of malicious sensor measurements that will go undetected.
The research in~\cite{ghafouri2018adversarial} presents an iterative algorithm for generating stealthy attacks on linear regression and feedforward neural network-based detectors.
The algorithm uses mixed-integer linear programming (MILP) to solve this problem.
For NN detectors, the algorithm first linearizes the network at each operating point and then solves the MILP problem.
The paper demonstrates a successful evasion attack on a simulated Tennessee Eastman process.

Recently, Erba \etal \cite{erba2019real} demonstrated a successful real-time evasion attack on an autoencoder-based detection mechanism in water distribution systems.
The authors of \cite{erba2019real} considered a white-box attacker that generates two different sets of spoofed sensor values: one is sent to the PLC, and the other is sent to the detector.

The most recent paper in this area \cite{zizzo2019intrusion} also focused on an adversary that can manipulate sensor readings sent to the detector.
The authors showed that such attackers can conceal most of the attacks present in the SWaT dataset.
Our study differs from these studies in a number of ways.
First and foremost, all of the abovementioned papers examined evasion attacks, while our research focuses on poisoning attacks.
Second, \cite{feng2017deep,erba2019real,zizzo2019intrusion} considered a threat model in which the attacker manipulates the detector's input data \textit{in addition} to manipulating the sensor data fed to the PLC.
Such a model provides a lot of freedom for the adversary to make changes to both types of data.
Our threat model considers a significantly more constrained attacker that can only change the sensory data that is provided \textbf{both} to the PLC and the detector.

A few recently published papers have studied poisoning attacks, however the authors considered them in a different context.
The study performed by Mu\~{n}oz-Gonz\'{a}lez \etal \cite{munoz2017towards} was the  first one to successfully demonstrate poisoning attacks on multiclass classification problems.
It also was the first to suggest generating poisoning data using back-gradient optimization.
Our research extends this method to semi-supervised multivariate time series regression tasks in the online training setting and evaluates the robustness of an autoencoder-based detector to such attacks.

Shafani \etal \cite{shafahi2018poison} and Suciu \etal \cite{suciu2018does} studied clean-label poisoning of classifiers.
In targeted clean-label poisoning, the attacker does not have control of the labels for the training data and changes the classifier's behavior for a specific test instance without degrading its overall performance for other test inputs.
These studies differ significantly from ours, both in terms of the learning task to be poisoned (classification vs. regression) and the domain (images vs. long interdependent multivariate time sequences).

Madani \etal~\cite{madani2018robustness} studied adversarial label contamination of autoencoder-based intrusion detection for network traffic monitoring.
Their research considered a black-box attacker that gradually adds \textit{existing} malicious samples to the training set, labeling them as normal.
Such a setting is very different from the one studied in our work.
First, we consider semi-supervised training; thus, there is no labeling involved.
Second, we explore algorithms for \textit{generating} adversarial poisoning samples that will direct the detector's outcome towards the target goal.

To summarize, although some of the previous research has dealt with related topics or domains, to the best of our knowledge, this study is the first one to address poisoning attacks on multivariate regression learning algorithms and specifically on online-trained physics-based anomaly and intrusion detectors in CPSs.

\section{Conclusions}
The reliability of NN-based cyber attack detectors depends on their resilience to adversarial data attacks.
In this study, we presented two algorithms for poisoning such detectors under a threat model relevant to online-trained ICSs.
The algorithms were evaluated on two datasets and found capable of poisoning the detectors both using artificial and real testbed data.
To the best of our knowledge, this is the first time adversarial poisoning of multivariate NN regression-based tasks with constraints is presented.

Our results also point out a number of factors influencing the robustness of the examined detectors to poisoning attacks.
First, in most of attacks the attacker had to generate long sequences of poisoning samples, often exceeding the amount of training data.
This creates practical difficulty in carrying out such attacks without being detected or affected by natural process changes.
Therefore, we can point out at a certain inherent robustness of the detectors.
In addition, the detectors can leverage the constraints of the NN  architecture used, to prevent the attacker from drifting the model to accept arbitrary values even with the unlimited attack time.
In our experiments, we used the \textit{tanh} activation in the last layer for such protection.

In addition, we evaluated three poisoning mitigation techniques: (1) decreasing the detection threshold, (2) increasing the sequence length, and (3) using dual detectors.
The mitigations were found effective and did not degrade the detection metrics significantly.

Some limitations and future directions of this study are worth noting.
First, this study only evaluated autoencoder-based detectors; studying the robustness of other NN architectures is a topic for future research.

This study could also be extended to evaluate the algorithms on more complex data involving attackers controlling multiple sensors.
Another important issue for future studies is increasing the efficiency of the poisoning algorithms proposed. 
Currently, they require significant computational time, thus making them inappropriate for real-time poisoning.
Lastly, in this study, we used a data-only approach for the algorithms' verification.
Validating the findings in a real-world testbed would be a natural next stage of this research, possibly after finding a way to accelerate poisoning sequence generation.

\subsection*{Acknowledgments}
This research was partially supported by the CONCORDIA project that has received funding from the European Union's Horizon 2020 research and innovation programme under grant agreement number 830927; by the PRIN 2017 project RexLearn (grant no. 2017TWNMH2), funded by the Italian Ministry of Education, University and Research; and by BMK, BMDW, and the Province of Upper Austria in the frame of the COMET Programme managed by FFG in the COMET Module S3AI.

\bibliographystyle{ACM-Reference-Format}
\bibliography{refs} 

\end{document}